\documentclass[12pt,draftcls,onecolumn]{IEEEtran} 

\usepackage{subfigure}
\usepackage{graphicx}
\usepackage{graphics} 
\usepackage{epsfig} 
\usepackage{mathptmx} 
\usepackage{times} 
\usepackage{amsmath} 
\usepackage{amssymb}  
\usepackage{amsthm}
\usepackage{dsfont}
\usepackage{color}

\newcommand{\beqnum}{\begin{equation}\begin{array}{lcl}}
\newcommand{\eeqnum}{\end{array}\end{equation}}
\newcommand{\beqnom}{\begin{eqnarray}}
\newcommand{\eeqnom}{\end{eqnarray}}
\newcommand{\beqnc}{\begin{center}\begin{eqnarray}}
\newcommand{\eeqnc}{\end{eqnarray}\end{center}}
\newcommand{\beqnlm}{\begin{equation}\vspace{-.5cm}\begin{array}{lll}}
\newcommand{\eeqnlm}{\end{array}\end{equation}}\vspace{-.5cm}
\newcommand{\beqay}{\begin{eqnarray*}}
\newcommand{\eeqay}{\end{eqnarray*}}
\newcommand{\bef}{\begin{figure}[tbh!]}
\newcommand{\enf}{\end{figure}}

\newcommand{\beq}{\begin{equation}}
\newcommand{\eeq}{\end{equation}}
\newcommand{\bed}{\begin{description}}
\newcommand{\eed}{\end{description}}
\newcommand{\lf}{\left\lfloor}
\newcommand{\rr}{\right\rceil}
\newcommand{\on}{\omega_n}
\newcommand{\oO}{\omega_0}

\newcommand{\sg}{\sigma}
\newcommand{\R}{\mathbb{R}}

\newtheorem{theorem}{\bf Theorem}
\newtheorem{rem}{\bf Remark}

\newtheorem{proposition}{\bf Proposition}

\newtheorem{Definition}{\bf Definition}

\title{$L_p$-stabilization of  integrator chains subject to input saturation using Lyapunov-based homogeneous design}
\author{Yacine Chitour, Mohamed Harmouche, Salah Laghrouche
\thanks{This research was partially supported by the iCODE Institute, research project of the Idex Paris-Saclay}
\thanks{Y. Chitour is with L2S, Universite Paris XI, CNRS 91192 Gif-sur-Yvette, France. (e-mail:yacine.chitour@lss.supelec.fr)}
\thanks{M. Harmouche is with Actility, Paris, France. (e-mail: mohamed.harmouche@actility.com)}
\thanks{S. Laghrouche is with OPERA Laboratory, UTBM, Belfort, France. (e-mail: salah.laghrouche@utbm.fr)}
}

\date{}

\begin{document}
\maketitle
{\bf Abstract} Consider the $n$-th integrator $\dot x=J_nx+\sigma(u)e_n$, where $x\in\R^n$, $u\in\R$, $J_n$ is the $n$-th Jordan block and $e_n=(0\ \cdots 0\ 1)^T\in\R^n$. We provide easily implementable state feedback laws $u=k(x)$ which not only render the closed-loop  system globally asymptotically stable but also are finite-gain $L_p$-stabilizing with arbitrarily small gain, as in \cite{Sab0}. 
These $L_p$-stabilizing state feedbacks are built from homogeneous feedbacks appearing in finite-time stabilization of linear systems. We also provide additional $L_\infty$-stabilization results for the case of both internal and external disturbances of the $n$-th integrator, namely for the perturbed system $\dot x=J_nx+e_n\sg (k(x)+d)+D$ where $d\in\mathbb{R}$ and $D\in\mathbb{R}^n$.

 \section{Introduction}
In this paper, we address robust stabilizability issues for an integrator chain subject to input saturation, i.e., System $(\Sigma)$
\beq\label{sys0}
(\Sigma)\qquad\dot x=J_nx+e_n\sigma(u),
\eeq
where $n$ is a positive integer, $x\in \R^n$, the matrix $J_n$ is the $n$-th Jordan block, i.e. the $n\times n$ matrix with entries $(J_n)_{ij}=1$ if $i=j-1$ and zero otherwise, the vector $e_n\in\R^n$ has all its coordinates equal to zero except the last one equal to one, and $\sg:\R\rightarrow\R$ is a saturation function whose prototype is the standard saturation function $\sg_0(s)=\frac{s}{\max (1,|s|)}$. 
In the sequel, we refer to System $(\Sigma)$ as the $n$-th integrator or an integrator chain of length $n$. Our purpose consists of investigating robustness properties associated with the (global asymptotic) stabilization to the origin of $(\Sigma)$. Note that semi-global stabilization issues  for linear systems subject to input saturation have been essentially all addressed, thanks to the work of Lin, Saberi and their coworkers by using ingenious low-and-high gain design technics (cf.  \cite{HuLin} and references therein). 

Consider then a \emph{stabilizing state feedback $k$ for $(\Sigma)$}, i.e., a static feedback law $u=k(x)$, where $k$ is a real-valued function defined on $\R^n$ so that every trajectory of the closed-loop system is globally asymptotically stable (GAS) with respect to the origin. Note that we do not assume $k$ to be even continuous, which will require if it is the case to precisely define solutions of Cauchy problems. 
Nevertheless, in order to test robustness of $k$, one considers, for $p\in [1,\infty]$,   the trajectories $x_d$ of the perturbed system 
\beq\label{sys-sat}
\dot x=J_nx+e_n\sg (k(x)+d),
\eeq
starting respectively from the origin if $p$ is finite and from any point of $\mathbb{R}^n$ if $p=\infty$ and which are associated to an arbitrary disturbance $d\in L_p(\R_+,\R)$, i. e. $d$ has finite $L_p$-norm ($\Vert d\Vert_p:=\Big(\int_{\R}\vert d(t)\vert^pdt\Big)^{1/p}<\infty$  if $p$ is finite and $\Vert d\Vert_\infty:=ess.\ supp. \ \vert d\vert<\infty$ if $p=\infty$).
Then, $k$ is said to be an \emph{$L_p$-stabilizing state feedback for $(\Sigma)$} if there exists $\gamma_p\in \mathcal{K}_{\infty}$ such that for every $d\in L_p(\mathbb{R}_+,\R)$ and $x_d$ defined as above, one has 
$\Vert x_d\Vert_p\leq \gamma_p(\Vert d\Vert_p)$ for $p$ finite and 
$\limsup_{t\rightarrow\infty}\Vert x_d(t)\Vert\leq \gamma_\infty( \Vert d\Vert_\infty)$ for 
$p=\infty$. The previous definition for $L_\infty$-stabilizability is called asymptotic gain property and it is required in the definition of Input to State Stability (ISS) introduced by Sontag, cf. \cite{Sontag-ISS}. In case the $ \mathcal{K}_{\infty}$ function $\gamma_p$ is linear, i.e., $
\gamma_p(x)=\gamma_p x$ for $x\geq 0$, the perturbed system is said to be \emph{finite-gain $L_p$-stable with finite gain $\gamma_p$}. 
One also says that Eq.~\eqref{sys-sat} stands for the $n$-th integrator subject to input saturation with \emph{internal disturbance} $d$ by opposition with the dynamics 
\beq\label{sys-sat-ext}
\dot x=J_nx+e_n\sg (k(x))+D, \quad D\in \mathbb{R}^n,
\eeq
which is referred as the $n$-th integrator subject to input saturation with \emph{external disturbance} $D$. 

The problem at stake belongs to a more general issue, that of stabilizing globally over $\mathbb{R}^n$ linear systems subject to input saturation of the type $(Sat)\ \dot x=Ax+B\sigma(u)$, where $x\in \R^n$, $u\in \mathbb{R}^p$ with $p$ a positive integer and the pair $(A,B)$ is controllable. Here, the $\mathbb{R}^p$-valued saturation function $\sigma(u)$  is equal to $(\sigma_1(u_1),\cdots,\sigma_p(u_p))^T$ where $u=(u_1,\cdots,u_p)$.

Global stabilization of $(Sat)$ can be achieved if and only if the eigenvalues of $A$ have non positive real part, cf. \cite{Sontag-book}. Most delicate issues arise when the spectrum of $A$ lies on the imaginary axis and we will assume that this is the case from the rest of the discussion. The first stabilizing state feedback $k_{opt}$ is the one given by the optimal control problem consisting of transferring any point of $\R^n$ to the origin in minimum time along trajectories of $(Sat)$, cf. \cite{ryan} for a description of the optimal synthesis corresponding to the double and triple integrators. However, it is immediate to see that, already for the double integrator, this feedback cannot ensure $L_p$-stability for any $p\in [1,\infty]$. Another candidate for stabilizing $(Sat)$ consists of taking linear state feedbacks $u=K^Tx$. In case $A$ is marginally stable (i.e., trajectories of $\dot x=Ax$ are bounded) or for $n$-th integrators with $n\leq 2$, one can find such linear stabilizing state feedbacks. As concerns their $L_p$-stabilization properties,
it was shown in \cite{LCS}  when $A$ is marginally stable that the linear state stabilizing feedback is also $L_p$-stabilizing for every $p\in[1,\infty]$, with additional results for external distubances. As for the double integrator, the linear stabilizing feedbacks are proved to be $L_p$-stabilizing for every $p\in[1,\infty]$ in \cite{chi01}, which also contains a partial answer for an open problem on $L_2$-stability proposed in \cite{blon99}:
that problem asks to compute the $L_2$-gain of the input-output map $d\mapsto \sigma(x+\dot x+d)$, i. e. the smallest positive number $\gamma_2$ such that for every disturbance $d\in L_2(\R_+,\R)$, one has $\Vert  \sigma(x+\dot x+d)\Vert_2\leq \gamma_2\Vert d\Vert_2$, where $x$ is the solution of the Cauchy problem $\ddot x=- \sigma(x+\dot x+d)$, $x(0)=\dot x(0)=0$. Besides the proof in \cite{chi01} that $\gamma_2$ is finite, 
non linear stabilizing state feedbacks with better performances than the linear ones (see also \cite{FGZ} for other non linear stabilizing state feedbacks) are also provided together with results for external distubances. One should notice that the robustness results of linear state feedbacks for the double integrator (and more generally planar systems) have been used for the robust stabilization of cascade and delay systems, cf. \cite{ACM,LCS2,YaChi1,YaChi2,GMM,MMS}.

It was then proved by Fuller and Sussmann, Yang (\cite{Fu,SuYa})
that the $n$-th integrator, $n\geq 3$ cannot be stabilized by linear state feedbacks $u=k^Tx$ and thus one has to resort to non linear state feedbacks. 
Thanks to Teel \cite{teel92} and Sussmann, Yang and Sontag \cite{SYS}, general and explicit stabilizing state feedbacks were constructed using nested saturations, i.e., feedbacks $N_l(\cdot)$
built inductively as follows: $N_0(x)=0$ and, for $1\leq j\leq l$, one sets 
$N_j(x)=\lambda_j\sigma_j(k_j^Tx+N_{j-1}(x))$ where the positive integer $l$ is the level of the nested satutation $N_l$, the $\lambda_j$'s are constants and the $k_j$'s are vectors of $\R^n$. However, by taking disturbances eventually equal to $d=-N_{p-1}(x)$ and using the abovementionned result of Fuller, Sussmann and Yang,  one readily deduces that nested saturations cannot be $L_p$-stabilizing feedbacks of the $n$-th integrator, $n\geq 3$ and $p\in [1,\infty]$. Related $L_2$-stabilization results for the feedbacks built with nested saturation  were obtained by Teel in \cite{teel96} for external disturbance $d$, i.e., for perturbed systems $\dot x=Ax+B\sg (k(x))+d$ where $(A,B)$ is controllable, the eigenvalues of $A$ have non positive real part and the disturbance $d$ has finite $L_2$-norm. One should also mention the construction of another type of stabilizing feedbacks due to Megretsky (cf. \cite{met96}), which are state dependant linear, i.e., of the type 
$u=B^TP(\varepsilon(x))x$, where the low-gain parameter $\varepsilon(x)$ is state-varying and defined as 
\begin{equation}\label{sche0}
\varepsilon(x)=\max\{r\in (0,1]\vert\ x^TP(r)x\  Tr(B^TP(r)B)\leq \Delta\},
\end{equation}
where $\Delta>0$ is fixed and $P(r)$ is the unique symmetric positive definite solution of a Ricatti equation parameterized by $r$. Then, using a variant of Megretsky feedbacks, Saberi, Hou and Stoorvogel were able to provide in \cite{Sab0}
the first solution to the finite-gain $L_p$-stabilisation problem associated to the internally perturbed system~\eqref{sys-sat} for $p\in [1,\infty]$.  
In addition, it has been recently shown in \cite{Saberi} that Megretsky feedbacks provide $L_\infty$-stabilization properties for  the $n$-th integrator subject to input saturation with external disturbances \eqref{sys-sat-ext}. In that work, no a priori bound only depending on the system is required for the external disturbance and more importantly a crucial distinction is pointed out between mismatched disturbance, i.e., $e_n^TD=0$ and matched disturbance, i.e., $e_j^TD=0$ for $1\leq i\leq n-1$, where the $e_i$'s are vectors in $\R^n$ with zero coordinates except the $i$-th one which is equal to one. However, the practical interest of these beautiful feedbacks is questionable.
Indeed the real-time implementation of that feedback requires the real-time solving of the optimization problem \eqref{sche0}. Furthermore, no approximated off-line computation can be envisioned based on finite covering of the state-space. 
To see that, first recall from \cite{Saberi} that the matrix $P(r)$ in Eq.~\eqref{sche0} is defined as the symmetric positive definite solution of $J_n^TP+PJ_n-Pe_ne_n^TP+rP=0$ and thus is equal to $rD_rP(1)D_r$ with $D_r=\hbox{diag}(r^{n-1},\cdots,r,1)$. Therefore, the mapping $r\mapsto P(r)$, defined on $(0,1]$ and taking values in the cone of real symmetric positive definite matrices  is strictly increasing 
as well the function $E_x(r)=r^2 x^TD_rP(1)D_rx$ defined for non zero $x$. It follows that the function $\varepsilon(\cdot)$ defined in Eq.~\eqref{sche0} is the unique solution in $(0,1]$ of $E_x(\varepsilon)=\Delta$ for non zero $x$. The fact that this equation is polynomial of degree $2n$ in $\varepsilon$ together with the fact that $lim_{\Vert x\Vert\rightarrow\infty}\varepsilon(x)=0$ (as shown in \cite{Sab0}) 
require that infinitely many quantized regions are necessary to cover the whole state-space in order to achieve off-line precomputation of \eqref{sche0}. This is why, eventhough \cite{Sab0} and \cite{Saberi} represent important breakthroughs, there is still need for easily implementable $L_p$-stabilizing feedbacks for perturbed systems \eqref{sys-sat} and \eqref{sys-sat-ext}.

In this paper, we provide yet another solution to the finite-gain $L_p$-stabilization of $(\Sigma)$ where our feedbacks are modifications of stabilizing feedbacks arising in the context of finite-time stabilization technics of the type $Lsign(\omega(x))$ for appropriate constant $L$ and continuous functions $\omega(\cdot)$, cf. \cite{Hong,HOSM1} and references therein. These feedbacks  are explicitely defined as Holder functions of the coordinates of the state $x$ and have been successfully implemented on practical examples of integrator chains, up to order four, cf. \cite{HOSM3,HOSM4,HOSM5}.

Trajectories of the corresponding closed-loop system $\dot x=J_n x+Le_nsign(\omega(x))$ converge to the origin in finite-time
and the crucial point lies in the fact that these feedbacks come together with global Lyapunov functions which are also ISS-Lyapunov for the perturbed system $\dot x=J_n x-Le_nsign(\omega(x)+d)$. To pass from these systems to systems given by 
Eq.~\eqref{sys-sat}, one has to replace the feedback $u=\omega(\cdot)$ in a neighborhood $\mathcal{V}$ of the origin by a linear feedback, which results in a global discontinuous feedback. The proof of the main result is then based on analytical manipulations using two positive definite functions, one being ISS-Lyapunov outside $\mathcal{V}$ and the other ISS-Lyapunov inside $\mathcal{V}$.  We finally extend these $L_p$-stabilization results for $L_\infty $-stabilization in the presence of both internal and external disturbances as in \cite{Saberi}. In particular, our feedbacks $L_\infty$-stabilize the perturbed system $\dot x=J_nx+e_n\sigma(u+d)+D$ where $D$ represents a mismatched external disturbance.

{\bf Acknowledgements.} The authors would like to thank A. Chaillet for constructive comments and suggestions.

\section{Notations and main definitions}
If $n$ is a positive integer, we consider for $1\leq i\leq n$ the vector $e_i\in \mathbb{R}^n$ having zero coordinates except the $i-$th one equal to $1$. We use $Id_n$ and $J_n$ respectively to denote the $n\times n$ identity matrix and the $n-$th Jordan block  respectively, the latter defined by $J_ne_i=e_{i-1}$ for $1\leq i\leq n$ with the convention that $e_j=0$ if $j\leq 0$ or $j>n$. If $A$ is any matrix, we use $A^T$ to denote the transpose of $A$. A function $\phi:\mathbb{R}_+\rightarrow \mathbb{R}_+$ is said to be of class $\mathcal{K}_{\infty}$ ($\phi\in \mathcal{K}_{\infty}$) if it is continuous, strictly increasing, $\phi(0)=0$ and $\lim_{s\rightarrow\infty}\phi(s)=\infty$. Recall that if $\phi\in \mathcal{K}_{\infty}$, then $\phi^{-1}\in \mathcal{K}_{\infty}$.

For $p\in [1,\infty)$ ($p=\infty$ respectively), we use $L_p(\mathbb{R}_+)$ ($L_\infty(\mathbb{R}_+)$ respectively) to denote the Banach space of measurable real-valued functions $f(\cdot)$ defined on $\mathbb{R}_+$ endowed with the $L_p$-norm $\Vert f\Vert_p:=\Big(\int_0^\infty\vert f(t)\vert^pdt\Big)^{1/p}$ ($\Vert f\Vert_\infty:=ess.\ supp. \ \vert f\vert$ respectively). 
If $K$ is a measurable set of $\mathbb{R}_+$ and $f\in L_p(\mathbb{R}_+)$ for finite $p$, we use $\vert K\vert $ and $\Vert f\Vert_{p,K}$ respectively to denote the Lebesgue mesure of $K$ and $\Big(\int_K\vert f(t)\vert^pdt\Big)^{1/p}$ respectively.
We define the function $sign$ as the multivalued function defined on $\mathbb{R}$ by $sign(x)=\frac x{\vert x\vert}$ for $x\neq 0$ and $sign(0)=[-1,1]$. Similarly, for every $a\geq 0$ and $x\in \mathbb{R}$, we use $\lf x\rr^a$ to denote $\left| x \right|^a sign(x)$. Note that $\lf \cdot\rr^a$ is a  continuous function for $a>0$ and of class $C^1$ with derivative equal to $a\left| \cdot \right|^{a-1}$ for $a\geq 1$. We use $s(\cdot)$ to denote the standard saturation function defined by $s(x)=\frac x{\max(1,\vert x\vert)}$ for $x\in \mathbb{R}$.

\begin{Definition}\label{def-sat}
An $S$-function (or saturation function) $\sigma:\mathbb{R}\rightarrow \mathbb{R}$ is any locally Lipschitz function so that 
\begin{description}
\item[$(i)$] there exists positive constants $a_1\leq a_2$ and $\frac{a_1}{b_1}\leq \frac{a_2}{b_2}$
for which the following inequality holds true for every $x\in \mathbb{R}$: 
$$
a_1x\ s(\frac x{b_1})\leq x\sigma(x)\leq a_2x\ s(\frac x{b_2});
$$
\item[$(ii)$] The limits $\sigma_{+\infty}:=\lim_{x\rightarrow +\infty}\sigma(x)$ and $\sigma_{-\infty}:=\lim_{x\rightarrow -\infty}\sigma(x)$ are defined, opposite and there exists a positive constant $C_\sigma$ such that, for $x\in \mathbb{R}$,
\beq\label{esti-SF}
\vert\sigma(\mid x\mid)-\sigma_{+\infty}\vert\leq \frac{C_\sigma}{1+\mid x\mid}.
\eeq
\end{description}
For $k>0$ and an $S$-function $\sigma(\cdot)$, we use $\sigma_k(\cdot)$ to denote the $S$-function $\sigma(k\cdot)$. For instance $s_k(\cdot)$, $\arctan_k(\cdot)$ and $\tanh_k(\cdot)$ are examples of $S$-functions for every $k>0$.
\end{Definition}
\begin{rem}\label{Rem0}
One can define a a saturation function only with Item $(i)$. It is for technical issues considered later in the paper that Item $(ii)$ is needed.
\end{rem}

In this paper, we consider stabilization issues for the control system $(\Sigma)$ defined in Eq.~\eqref{sys0}, where $n$ is a positive integer, $x\in \mathbb{R}^n$, $u\in \mathbb{R}^n$ and $\sigma$ is an $S$-function. This is essentially equivalent as considering the control system on $\mathbb{R}^n$ 
given by $\dot x=J_nx+e_nu$, with bounded control $u$. Notice that the bound on the amplitude of $u$ is irrerelevant as regards feedback stabilization since multiplying  $\dot x=J_nx+e_nu$ by a positive constant $C$ and making the linear change of variable $y=Cx$ only changes the bound on the amplitude of $u$.

We next provide the definition of a stabilizing feedback for $(\Sigma)$.
\begin{Definition}\label{def-SF}
We say that the function $k:\mathbb{R}^n\rightarrow\mathbb{R}$ is a stabilizing feedback (SF) for $(\Sigma)$ if the closed-loop system $\dot x=J_nx+e_n\sigma(k(x))$ is globally asymptotically stable (GAS) with respect to the origin. Note that $k$ can possibly be discontinuous so in the case where $k$ is not locally Lipschitz, one must not only define specifically what the solutions of Cauchy problems are and guarantee that the origin is GAS with respect to all of them. 
\end{Definition}
We next provide a notion of robustness of a stabilizing feedback (see )which generalizes that of linear systems, cf \cite{Sontag-book}.
\begin{Definition}\label{def-LpSF}
Let $p\in [1,\infty]$. We say that the function $k:\mathbb{R}^n\rightarrow\mathbb{R}$ is an $L_p$-stabilizing feedback ($L_p$-SF) for $(\Sigma)$ if there exists $\gamma_p\in \mathcal{K}_{\infty}$ such that for every $d\in L_p(\mathbb{R}_+)$ and $x_d$ in the set of trajectories of 
\beq\label{Lp-stab-S}
\dot x=J_nx+e_n\sigma(k(x)+d),\quad 
\left\{\begin{array}{lll}&x(0)=0\ \ \hbox { for $p$ finite, }\\
&x(0)\in \mathbb{R}^n \hbox { for $p=\infty$,}
\end{array}\right.
\eeq
one has 
\bed
\item[$(L_p-S)$] $\quad$ $ \Vert x_d\Vert_p\leq \gamma_p(\Vert d\Vert_p)$ for $p$ finite;
\item[$(L_\infty-S)$]  $\quad$ $\limsup_{s\rightarrow\infty} \Vert x_d(s)\Vert\leq \gamma_\infty(\Vert d\Vert_\infty)$.  Sometimes one can consider another statement where the left hand-side of the previous inequality is replaced by $\Vert x_d\Vert_\infty$ while assuming that the trajectory starts at the origin.
\eed
The function $\gamma_p\in \mathcal{K}_{\infty}$ is referred as the gain function. When it is linear, i.e., $\gamma_p(x)=\gamma_px$ for $x\geq 0$, then $(\Sigma)$ is said to be finite-gain $L_p$-stabilizable by $u=k(x)$ with finite gain $\gamma_p$.
\end{Definition}
\begin{rem}\label{Rem1}
If $(\Sigma)$ admits an $L_p$-stabilizing feedback $k(\cdot)$ for some $p\in [1,\infty)$, then
$k(\cdot)$ is also a stabilizing feedback for $(\Sigma)$. This is essentially established in Item$(1)$ of \cite[Lemma 5]{LCS}.
\end{rem}
\begin{rem}\label{Rem3}
Assume that $k:\mathbb{R}^n\rightarrow\mathbb{R}$ is an $L_p$-stabilizing feedback ($L_p$-SF) for $(\Sigma)$ for some $p\in [1,\infty)$. From Items $(1)$ and $(3)$ of  \cite[Lemma 4]{LCS}, one gets that, for every $d\in L_p(\mathbb{R}_+)$, any $x_d$ in the set of solutions of Eq.~\eqref{Lp-stab-S} tends to zero as $t$ tends to infinity. If moreover, $k$ is differentiable at zero and $J_n+\sigma(0)e_nK^T$ is Hurwitz with $K:=\nabla k(0)$, then for every solution $x_d$ of Eq.~\eqref{Lp-stab-S} belongs to $L_p(\mathbb{R}_+)$.
\end{rem}

\section{Preliminary solution to the $L_p$-stabilization problem}
As mentionned in Introduction, the purpose of this paper consists in constructing an $L_p$-stabilizing feedbacks ($L_p$-SF) for $(\Sigma)$ for every $p\in [1,\infty]$. To proceed, we actually start with a preliminary solution for the $L_p$-stabilization of $(\Sigma)$ where the saturation function is replaced by the function $sign$. More precisely, we consider the stabilization of $(\Sigma)$ given in \eqref{sys0} by the feedback $-l_nsign(\omega_n(x))$ where $l_n$ is a positive constant (to be defined) and the feedback law $\omega_n(\cdot)$ defined inductively as follows (cf. \cite{Hong} and references therein). 

Define the following parameters:
\beq\label{param0}
p_i = 1-\frac{i-1}n,\ \ 1\leq i\leq n+1\hbox{ and }\beta_0 = p_2,\ \beta_i  =\frac{n-1+i}{n-i}  ,\ 1\leq i\leq n-1.
\eeq
Note that $p_{n+1}=0$, $\beta_0<1$ and $\beta_i>1$ for $1\leq i\leq n$. Then, given positive constants $l_i$, $1\leq i\leq n$, define the following functions for $0\leq i\leq n$
\beqnum \label{u}
	\left\{
		\begin{array}{ccl}
			v_0 &\equiv& 0,\\
			v_i(x_1,\cdots,x_i) &=& -l_i \lf  \omega_i(x_1,\cdots,x_i)\rr^{ \frac{p_{i+1}}{p_i \beta_{i-1}} },\ \omega_i=\lf x_i \rr^{\beta_{i-1}}-\lf v_{i-1}(x_1,\cdots,x_{i-1}) \rr^{\beta_{i-1}}.
		\end{array}
	\right.
\eeqnum
Note that $v_i$ is defined on $\mathbb{R}^i$ for $1\leq i\leq n$ and $v_n(x)=-l_n sign(\omega_n(x))$. 
One has then the following theorem.
\begin{theorem}\label{th0}(\cite{Hong})
There exists positive constants  $l_i$, $1\leq i\leq n$, such that the controller $u=v_n(x)$ is a stabilizing feedback for the control system $\dot x=J_nx+e_nu$, with $\vert u\vert\leq l_n$.
Moreover, this stabilization occurs in finite time.
\end{theorem}
Since the the feedback law $u=v_n(x)$ is discontinous, solutions of Cauchy problem must be specified. Here, solutions correspond to Filippov solutions (see \cite{Filippov} for a definition of such solutions) associated to the differential inclusion $\dot x\in J_nx-l_ne_nsign(\omega_n(x))$. 
This fundamental result is obtained by building a Lyapunov function which will be instrumental for the rest of the paper. We provide its construction below. 
For $1\leq i\leq n$, first define $W_i:\mathbb{R}^i\rightarrow \mathbb{R}_+$ as 
\begin{equation}\label{wi-i}
W_i(x_1,\cdots,x_i) = \int_{v_{i-1}}^{x_i}\lf s\rr^{\beta_{j-1}}-\lf v_{i-1} \rr^{\beta_{i-1}}ds=
\frac{ \left| x_i \right|^{\beta_{i-1} + 1} - \left| v_{i-1}  \right|^{\beta_{i-1} + 1}}{\beta_{i-1} + 1} - \lf
v_{i-1}\rr^{\beta_{i-1}} \left(  x_i- v_{i-1} \right).
\end{equation}
Note that $\frac{\partial{W_i}}{\partial{x_i}}=\omega_i(x_1,\cdots,x_i)$. Then the Lyapunov function $V_n$ is defined as 
\beq\label{V_n}
V_n(x)=\sum_{i=1}^nW_i(x_1,\cdots,x_i),
\eeq
and one has $\frac{\partial{V_n}}{\partial{x_n}}=\omega_n(x)$. The key inequality then is the following one. Thanks to homogeneity properties, the time derivative of $V_n$ along non trivial trajectories of $\dot x=J_nx+e_nu$, which is denoted by $\dot V_n$, can be upper bounded by
\beq\label{der0}
\dot V_n\leq -c_nV_n^{\alpha}(x)+\omega_n(x)(u+l_nsign(\omega_n(x)),
\eeq
where $c_n$ is a postive constant and $\alpha:=\frac{2(n-1)}{2n-1}<1$. If one chooses the feedback law $u=-l_nsign(\omega_n(x))$, Theorem~\ref{th0} follows at once.

\begin{rem}
In \cite{Hong}, Theorem \ref{th0} is established for homogeneity degrees $(-1/n,0)$ only. However, the proof there extends readily to the case of a homogeneity degree equal to $-1/n$
which corresponds to what is given in the present paper, as well as to the case of a homogeneity degree equal to zero, which corresponds to a linear feedback. 
\end{rem}

Note also the following technical inequality (to be used later) holds true: for every $C>0$, there exists $K(C)>0$ such that, along any trajectory $x(\cdot)$ of $\dot x=J_nx+e_nu$ with $\vert u\vert\leq 1$, the time derivative $ \dot V_n$ of $V_n(x(\cdot))$ verifies a. e.
\beq\label{tech0}
\vert \dot V_n\vert\leq K(C)V_n^{\alpha}(x(t)),\hbox{ if }V_n(x(t))\geq C.
\eeq
To be completely rigorous, Eq.~\eqref{der0} actually holds almost everytwhere on the open set of times $t$ so that $x(t)\neq 0$. For $L_p$-stabilization purposes, one can always work on this set of times. We will therefore assume for the rest of the paper and without further mention that we evaluate quantities of interest along pieces of non trivial trajectories passing through the origin at isolated times.  

We now proceed with the $L_p$-stabilization of the control system $\dot x=J_nx+e_nu$. However, we must consider a similar definition to that given in Definition~\ref{def-LpSF} where the S-function $\sigma$ is replaced by the function $sign$. We then consider the trajectories of the perturbed system
\beq\label{Lp-stab-sign}
\dot x=J_nx-l_ne_nsign(\omega_n(x)+d),\quad 
\left\{\begin{array}{lll}&x(0)=0\ \ \hbox { for $p$ finite, }\\
&x(0)\in \mathbb{R}^n \hbox { for $p=\infty$,}
\end{array}\right.
\eeq
where $d\in L_p(\mathbb{R}_+)$ and $p\in[1,\infty]$.

We prove the following result, which is reminiscent of $L_p$-stabilization.
\begin{theorem}\label{th1}
Let $p\in [1,\infty]$. For every $d\in L_p(\mathbb{R}_+)$ and $x_d$ in the set of solutions of the Cauchy problem defined by Eq.~\eqref{Lp-stab-sign}, one has
\bed
\item[$(sign)_p$] $\quad$ $\Vert V_n^{\alpha}(x_d)\Vert_p\leq \frac{2l_n}{c_n}\Vert d\Vert_p$ for $p$ finite. Moreover, if $\beta:=\alpha(p-1)$, one has that $$\Vert V_n(x_d)\Vert_{\infty}\leq \big(\frac{(2l_n)^p(1+\beta)}{c_n^{p-1}}\big)^{\frac1{1+\beta}}\Vert d\Vert_p^{\frac{p}{1+\beta}},$$ and $x_d$ tends to zero at infinity;
\item[$(sign)_\infty$]  $\quad$ $\limsup_{s\rightarrow\infty}V_n^{\alpha}(x_d(s))\leq \frac{2l_n}{c_n}\Vert d\Vert_\infty$.
\eed
\end{theorem}
{\bf Proof.} The key inequality relative to Eq.~\eqref{Lp-stab-sign} is the following. For every measurable function $d$ defined on $\mathbb{R}_+$ and every non trivial trajectory of Eq.~\eqref{Lp-stab-sign}, the time derivative of $V_n$ along such a trajectory verifies, for almost every non negative time,
\beq\label{der1}
\dot V_n(t)\leq -c_nV_n^{\alpha}(x(t))+2l_n\vert d(t)\vert.
\eeq
Indeed, from Eq.~\eqref{der0}, one deduces that 
$$
\dot V_n(t)\leq -c_nV_n^{\alpha}(x(t))+l_n\omega_n(x(t))\big(sign(\omega_n(x(t)))-sign(\omega_n(x(t))+d(t))\big).
$$
If $\vert \omega_n(x(t))\vert >\vert d(t)\vert$, then $sign(\omega_n(x(t)))=sign(\omega_n(x(t))+d(t))$ and if  $\vert \omega_n(x(t))\vert \leq\vert d(t)\vert$, then 
$$
\vert \omega_n(x(t))\big(sign(\omega_n(x(t)))-sign(\omega_n(x(t))+d(t))\big)\vert \leq 2\vert d(t)\vert.
$$
From Eq.~\eqref{der1}, we deduce at once Item $(sign)_\infty$. 

As regards Item $(sign)_p$ for $p\in[1,\infty)$, set $\beta=\alpha(p-1)$. We first multiply Eq.~\eqref{der1} by $V_n^{\beta}(x(t))$ and then integrate it between $t=0$ and $t=T$ where $T>0$ is arbitrary. We obtain that
\begin{equation}\label{good0}
\frac{V_n^{\beta+1}(x(T))}{\beta+1}+c_n\int_0^TV_n^{\alpha p}(x(t))dt\leq 2l_n\int_0^T\vert d(t)\vert V_n^{\beta}(x(t))dt.
\end{equation}
If $p=1$, we immediately obtain the inequality in Item $(sign)_1$ by letting $T$ tend to infinity. If $p>1$, we apply Holder's inequality to the right-hand side of the above inequality and proceed as for $p=1$ to get  the first inequality in Item $(sign)_p$.  

For the sup-norm estimate, one plugs the $L_p$ estimate of $V_n^{\alpha}$ to get that, for every 
$T\geq 0$, 
$$
\frac{V_n^{\beta+1}(x(T))}{\beta+1}\leq 2l_n\Vert d\Vert_p\Vert V_n^{\alpha}\Vert_p^{p-1},
$$
thus implying the second part of Item $(sign)_p$.

To obtain the claim on convergence to zero as time tends to infinity, we first notice that 
$\liminf_{t\rightarrow\infty}V_n(x(t))=0$ due to the convergence of the integral. Reasoning by contradiction, we deduce the existence of $\varepsilon>0$ and two sequences of times $(s_l)$
and $(t_l)$ such that, for $l\geq 1$, 
$$
s_l<t_l,\ \lim_{l\rightarrow\infty}s_l=\infty,\ \lim_{l\rightarrow\infty}V_n(x(s_l))=0,\ V_n^{\beta+1}(x(t_l))\geq \varepsilon.
$$
Multiplying Eq.~\eqref{der1} by $V_n(x(t))^{\beta}$ and then integrate it between $t=s_l$ and $t=t_l$ , we obtain that
$$
\varepsilon\leq V_n^{\beta+1}(x(t_l))\leq V_n^{\beta+1}(x(s_l))+ 2l_n(1+\beta)\int_{s_l}^{t_l}\vert d(t)\vert V_n^{\beta}(x(t))dt.
$$
Since the right-hand side converges to zero as $l$ tends to infinity, we derive a contradiction and conclude the proof of the theorem.

\begin{rem}\label{sgnISS}
The differential inequality \eqref{der1} shows that $V_n$ is an ISS-Lyapunov function for 
$\dot x=J_nx-l_ne_nsign(\omega_n(x)+d)$, rendering that system ISS according to \cite[Theorem 5]{Sontag-ISS}
\end{rem}

\section{Solution to the finite-gain $L_p$-stabilization problem}
First of all, one can use $u=sign(\omega_n(x))$ to stabilize $\dot x=J_n x-\frac{l_n}{\sigma_\infty} e_n\sigma(u)$ but this feedback is not an $L_p$ stabilizing feedback for any $p\in [1,\infty]$ since the perturbation $d=-sign(\omega_n(x))$ after a certain time on appropriate intervals of time would yield arbitrarily large trajectories. The second attempt woud consist in taking $u=\omega_n(x)$. 
We are not able to prove that it is a stabilizing feedback for $(\Sigma)$, i.e., the closed-loop system $\dot x=J_nx-l_ne_n\sigma(\omega_n(x))$ is GAS with respect to the origin. We however get the following proposition.

\begin{proposition}\label{partial0}
Consider the perturbed system
$\dot x=J_nx-l_ne_n\sigma_k(\omega_n(x)+d)$ where $\sigma$ is an $S$-function, $k>0$ and $d\in L_\infty(\mathbb{R}_+$. Then, there exists a positive constant $C>0$ and $k$ large enough such that, along any non trivial trajectory of the above perturbed system, one gets 
\begin{equation}\label{prop-partial}
\limsup_{s\rightarrow\infty}V_n^{\alpha}(x_d(s))\leq \frac{2l_n}{c_n}(\frac{1+C_\sigma}k+2\Vert d\Vert_\infty).
\end{equation}
\end{proposition}
{\bf Proof.} This simply results from Eq.~\eqref{partial1}.

Moreover, numerical simulations (with $\sigma=s_k$, $k>0$ large) seem indicating that it does not hold true. Indeed, the problem occurs when trajectories appproach the origin, and in that case, the saturated feedback $\sigma(\omega_n(\cdot))$ tends to zero (instead of keeping a constant amplitude as compared to the feedback $sign(\omega_n(x))$) loses its stabilizing effect. This is why we had to replace the feedback $u=\omega_n(x)$ in a neighborhood of the origin, obtaining a discontinuous feedback. 

For that purpose, we consider $K\in\mathbb{R}^n$ and a real symmetric positive matrix $P$ such that, for every $\rho\in [\frac{a_1}{b_1},\frac{a_2}{b_2}]$, it holds
$$
(J_n-\rho l_ne_n K^T)^TP+P(J_n-\rho l_ne_nK^T)\leq-Id_n.
$$
Such $K$ and $P$ do exist according to \cite{CS} (which was inspired by \cite{GK}). For $x\in\mathbb{R}^n$, define the positive definite function $V_0(x)=(x^TPx)^{1/2}$ and the feedback $\omega_0(x)=K^Tx$. Note that one has the following inequality along every non trivial trajectory of $\dot x=(J_n-r(t)l_ne_nK^T)x+e_nd$,
\beq\label{der2}
\dot V_0\leq -c_0V_0+l_0\vert d\vert,
\eeq
where $c_0,l_0$ are positive constants and $r(\cdot)$ is any measurable function taking values in $[\frac{a_1}{b_1},\frac{a_2}{b_2}]$.

For $k>0$, we then define the feedback $\omega:\mathbb{R}^n\rightarrow\mathbb{R}$ by
\beq\label{feed0}
\omega(x)=\left\{\begin{array}{ccc}\on(x),&\hbox{ if }V_0(x)>A,\\
\frac{\oO(x)}k,&\hbox{ if }V_0(x)\leq A,\end{array}\right.
\eeq
where the constant $A$ is chosen small enough so that 
\beq\label{estA}
\max_{V_0(x)\leq A}\vert\oO(x)\vert\leq \min(1,b_1,b_2).
\eeq 

We next state the main result of the paper.

\begin{theorem}\label{Th-p} For $A>0$ small enough so that Eq.~\eqref{estA} holds true, $\sigma$ an $S$-function and $k>0$ large enough, System $(\Sigma)$ given by $\dot x=J_nx+e_n\sigma(u)$ is finite-gain $L_p$-stabilizable by the state feedback $u=k\omega(\cdot)$ for every $p\in [1,\infty]$.
\end{theorem}

\begin{rem}\label{attenuation} One must recall that the fundamental work \cite{Sab0} provides a finite-gain $L_p$-stabilizer with arbitrarily small gain. In our case we reach the same conclusion by simply reparameterizing the trajectories of $\dot x=J_nx-l_ne_n\sigma(\omega_n(x))$ to $rD_rx(\frac{\cdot}r)$, where $r>0$ and $D_r=\hbox{diag}(r^{n-1},\cdots,r,1)$.
\end{rem}

The proof of Theorem~\ref{Th-p} is actually based on the next proposition. To state it, we need the following definition. Let $W$ be the positive definite function over $\mathbb{R}^n$ defined by $W(x)=\min(V_0(x),V_n^\alpha(x))$ which tends to infinity as $\Vert x\Vert$ tends to infinity.

\begin{proposition}\label{th3}
For $A>0$ small enough so that Eq.~\eqref{estA} holds true, $\sigma$ an $S$-function and $k>0$ large enough, the feedback $k\omega(\cdot)$ defined in Eq.~\eqref{feed0} is an $L_p$-stabilizing feedback for $\dot x=J_nx-\frac{l_n}{\sigma_{+\infty}}e_n\sigma(u)$ for every $p\in[1,\infty]$. More precisely, we prove that, for $A>0$ small enough so that Eq.~\eqref{estA} holds true, $\sigma$ an $S$-function and $k>0$ large enough,
\begin{description}
\item[$(S-\infty)$] $ $ if $p=\infty$, there exists $C_\infty>0$ such that, for every $d\in L_\infty(\mathbb{R}_+)$ and trajectory of $\dot x=J_nx-\frac{l_n}{\sigma_{+\infty}}e_n\sigma(k\omega(x)+d))$, one has
\beq\label{th-inf}
\limsup_{s\rightarrow\infty}W(x(s))\leq C_\infty\Vert d\Vert_\infty.
\eeq
\item[$(S-p)\ \ $] $ $  If $p\in [1,\infty)$, there exists $C_p>0$ such that, for every $d\in L_p(\mathbb{R}_+)$, one has 
\beq\label{th-fin}
\Vert W(x(\cdot))\Vert_p\leq C_p\Vert d\Vert_p,
\eeq
for every trajectory of $\dot x=J_nx-\frac{l_n}{\sigma_{+\infty}}e_n\sigma(k\omega(x)+d))$ starting at the origin and all of them converge to the origin at infinity.
\end{description}
\end{proposition}

{\bf Proof of Proposition~\ref{th3}. } Up to a linear change of variable, we assume with no loss of generality that $\sigma_{+\infty}=1$. We also fix $A$ small enough so that Eq.~\eqref{estA} holds true.

We first set some notations. We use $V_{0,>}^A$, $V_{0,\leq}^A$, $V_{0,<}^A$ and $V_{0,=}^A$
respectively to denote the sets $\{x\mid V_0(x)>A\}$, $\{x\mid V_0(x)\leq A\}$, $\{x\mid V_0(x)< A\}$ and $\{x\mid V_0(x)=A\}$ respectively. For $T\geq 0$, we set $V_{0,>}^{A,T}$, $V_{0,\leq}^{A,T}$ and $V_{0,=}^{A,T}$ respectively as the intersections of $V_{0,>}^A$, $V_{0,\leq}^A$, $V_{0,<}^A$ and $V_{0,=}^A$ with $[0,T]$ respectively. Finally set $v_A=\min_{x\in V_{0,=}^A}V_n(x)$ and $V_A=\max_{x\in V_{0,=}^A}V_n(x)$.

Since we are dealing with a discontinuous feedback, we must precise what we mean by solutions of $\dot x=J_nx-l_ne_n\sigma(k\omega(x)+d)$. It is enough to consider the case $d=0$. First, define for $x\in\mathbb{R}^n$ the closed interval $I(x)$ of $\mathbb{R}$ delimited by $\sigma_k(\on(x))$ and $\oO(x)$. In the open set $V_{0,>}^A$, trajectories are absolutely continuous curves solutions of a differential equation with continuous right hand-side. At its boundary $V_{0,=}^A$, the selection made among trajectories of the differential inclusion 
$\dot x\in J_nx-l_ne_nI(x)$ as given by Eq.~\eqref{feed0} is well-defined because any nontrivial trajectory of $\dot x=J_nx-l_ne_n\sigma(K^Tx)$ starting on $V_{0,=}^A$ stays in $V_{0,\leq}^A$ for all non negative times. 

The proof of the theorem is based on the following two inequalities whose proofs are given in Appendix.
\begin{description}
\item[$(i)$] On the open set $V_{0,>}^A$, the time derivative $\dot V_n(\cdot)$ of $V_n$ along trajectories of $\dot x=J_nx-l_ne_n\sigma(k\omega(x)+d)$ verifies almost everywhere
\beq\label{ineq1}
\dot V_n\leq -\frac{c_n}2V_n^{\alpha}(x(t))+4l_n\vert d\vert.
\eeq
\item[$(ii)$] On the closed set $V_{0,\leq }^A$, the time derivative $\dot V_0(\cdot)$ of $V_0$ along non trivial trajectories of $\dot x=J_nx-l_ne_n\sigma(k\omega(x)+d)$ verifies almost everywhere 
\beq\label{ineq2}
\dot V_0\leq -\frac{c_0}2V_0(x(t))+4l_0\min(1,\vert d\vert).
\eeq
\end{description}
We start with the case $p=\infty$. Let $x(\cdot)$ be a  non trivial trajectory of $\dot x=J_nx-l_ne_n\sigma(k\omega(x)+d)$. 

Assume first that there exists $t_0\geq 0$ such that  one of the following alternatives occurs: 
\begin{description} 
\item[$(a)$] either $V_0(x(t))\leq A$ for every $t\geq t_0$, and then $\limsup_{s\rightarrow\infty}V_0(x(s))\leq \frac{8l_0}{c_0}\Vert d\Vert_\infty$ by using Eq.~\eqref{ineq2};
\item[$(b)$] or $V_0(x(t))> A$ for every $t\geq t_0$, and then $\limsup_{s\rightarrow\infty}V_n^{\alpha}(x(s))\leq \frac{8l_n}{c_n}\Vert d\Vert_\infty$ by using Eq.~\eqref{ineq1}.
\end{description}
If such a $t_0$ does not exist, then one has $V_{0,>}^A=\cup_{k\geq 0}I_k$ where $I_k=(s_k,t_k)$ is a non-empty interval, $\lim_{k\rightarrow\infty}s_k=\infty$ and there is a subsequence $(k_l)$ tending to infinity so that $t_{k_l}<s_{k_l+1}$. By integrating 
Eq.~\eqref{ineq2}  on $[t_{k_l},s_{k_l+1}]$ (or part of it), one gets that $A\leq \frac{16l_0}{c_0}\Vert d\Vert_\infty$. Set $L:=\limsup_{s\rightarrow\infty}V_n^{\alpha}(x(s))$. If $L\leq 2V_A^\alpha$, then $L\leq C_2\Vert d\Vert_\infty$
with $C_2=\frac{32V_A^\alpha l_0}{Ac_0}$. If $L> 2V_A^\alpha$, there exists, for $\varepsilon>0$ small enough and up to a subsequence, $\tilde{s}_k<\tilde{t}_k$ in $I_k$ for every $k\geq 0$ so that,
$$V_n^{\alpha}(x(\tilde{s}_k))=V_n^{\alpha}(x(\tilde{t}_k))=L-\varepsilon,\hbox{ and }
V_n^{\alpha}(x(s))>L-\varepsilon\hbox{ on }(\tilde{s}_k,\tilde{t}_k).
$$
Integrating Eq.~\eqref{ineq1} on $[\tilde{s}_k,\tilde{t}_k]$, then letting $\varepsilon$ tend to zero, one gets $(b)$. That concludes the proof of Item $(S-\infty)$.

We next turn to the proof of the theorem for $p\in [1,\infty)$. Let $x(\cdot)$ be a  non trivial trajectory of $\dot x=J_nx-l_ne_n\sigma(k\bar{\omega}(x)+d)$. 
For $T>0$, one has the following disjoint union
$$
[0,T]=V_{0,>}^{A,T}\cup V_{0,<}^{A,T}\cup V_{0,=}^{A,T}.
$$ 

Assume first that $V_{0,>}^{A,T}$ is empty. By multiplying Eq.~\eqref{ineq2} by $V_0^{p-1}$ and integrating over $[0,T]$, one gets that 
$$
\Vert V_0\Vert_{p,[0,T]}\leq  \frac{8l_0}{c_0}\Vert d\Vert_{p,[0,T]}.
$$
Assume now $V_{0,>}^{A,T}$ is non empty and thus $V_{0,=}^{A,T}$ is non empty as well.

Multiplying Eq.~\eqref{ineq1} by $V_n^{\alpha(p-1)}$, integrating it over $V_{0,>}^{A,T}$ and applying Holder's inequality if $p>1$ leads to 
$$
\int_{V_{0,>}^{A,T}}V_n^{\alpha(p-1)}\dot V_n+\frac{c_n}2\int_{V_{0,>}^{A,T}}V_n^{\alpha p}(x(t))dt\leq 4l_n\int_{V_{0,>}^{A,T}}V_n^{\alpha(p-1)}\vert d\vert\leq 4l_n \Vert d\Vert_{p,V_{0,>}^{A,T}}
\Vert V_n^{\alpha p}\Vert^{p-1}_{p,V_{0,>}^{A,T}}.
$$
By applying now Young's inequality if $p>1$ to the right-hand side of the above set of inequalities, one deduces that there exists a positive constant $C_{1,p}$ only depending on $c_n,l_n$ and $p$
so that 
\beq\label{eq>}
\int_{V_{0,>}^{A,T}}V_n^{\alpha(p-1)}\dot V_n+\frac{c_n}4\int_{V_{0,>}^{A,T}}V_n^{\alpha p}(x(t))dt\leq C_{1,p}\Vert d\Vert^p_{p,V_{0,>}^{A,T}}.
\eeq
The absolutely continuous function $t\mapsto V_0(x(t))$ is constant on the measurable set $V_{0,=}^{A,T}$. If its Lebesque measure $\mid V_{0,=}^{A,T}\mid$  is positive, then there exists $F\subset V_{0,=}^{A,T}$ with $\mid F\mid=\mid V_{0,=}^{A,T}\mid$ so that the time derivative of $V_0(x(t))$ is equal to zero for $t\in F$. By using Eq.~\eqref{ineq2}, we get that, for almost every $t\in V_{0,=}^{A,T}$, $A=V_0(x(t))\leq  \frac{8l_0}{c_0}\vert d(t)\vert$. That implies that $A\mid V_{0,=}^{A,T}\mid^{1/p}\leq \Vert d\Vert_{p,V_{0,=}^{A,T}}$. On the other hand, integrating Eq.~\eqref{tech0} over $V_{0,=}^{A,T}$ yields that
$$
\int_{V_{0,=}^{A,T}}V_n^{\alpha(p-1)}\vert\dot V_n\vert\leq K(v_A)V_A^{\alpha(p-1)}\int_{V_{0,=}^{A,T}}V_n^{\alpha}\leq K(v_A)V_A^{\alpha p} \mid V_{0,=}^{A,T}\mid\leq \frac{K(v_A)V_A^{\alpha p}}{A^p}  \Vert d\Vert^p_{p,V_{0,=}^{A,T}}.
$$
By using Young's inequality if $p>1$, we deduce that there exists a positive constant $C_{2,p}$ only depending on $c_n,l_n$ and $p$
such that 
\beq\label{eq=}
\int_{V_{0,=}^{A,T}}V_n^{\alpha(p-1)}\vert\dot V_n\vert+\frac{c_n}4\int_{V_{0,=}^{A,T}}V_0^p(x(t))dt\leq C_{2,p}\Vert d\Vert^p_{p,V_{0,=}^{A,T}}.
\eeq
It remains to obtain a similar estimate on $V_{0,<}^{A,T}$. The latter is an open set of $[0,T]$
and since the trajectory starts at the origin, one has that $V_{0,<}^{A,T}=\cup_{0\leq j\leq J}I_j(s_j,t_j)\cup I_f$, where $J\leq \infty$, $I_0=[s_0,t_0)$ with $s_0=0$, $I_j=(s_j,t_j)$ for $1\leq j\leq J$ and $I_f$ is either empty or equal to $(s_f,t_f]$ with $t_f=T$. Then $V_0(x(t))=A$ for $t=t_0,s_f$ and $t=s_j,t_j$ for $1\leq j\leq J$. One next multiplies Eq.~\eqref{ineq2} by $V_0^{p-1}$, integrate it and apply Holder inequality if $p>1$ on each interval $I_j$, $0\leq j\leq J$ and on $I_f$. One then obtains
\beq\label{=00}
E+\frac{c_0}2\Vert V_0\Vert^p_{p,I}\leq 4l_0\int_{I}V_0^{p-1}\vert d\vert\leq 4l_0\Vert d\Vert_{p,I}\Vert V_0\Vert^{p-1}_{p,I},
\eeq
where $E=\frac{A^p}p$ if $I=I_0$, $E=0$ if $I=I_j$, $1\leq j\leq J$ and $E=\frac{V_0(x(T))^p-A^p}p$ if $I=I_f$. By using Young's inequality if $p>1$, we deduce that there exists a positive constant $C_{3,p}$ only depending on $c_n,l_n$ and $p$ such that 
\beq\label{=0j}
E+\frac{c_0}4\int_{I}V_0^p\leq C_{3,p}\Vert d\Vert^p_{p,I},
\eeq
with the same notational conventions for $E,I$ as above. 

We now need to upper bound $Int_I:=\int_IV_n^{\alpha(p-1)}\dot V_n$ by a constant times $\Vert d\Vert^p_{p,I}$ on each interval $I$. 
For $I=I_0$, setting $C_{4,p}=\frac{V_A^{\alpha p}}{\alpha A^p}$, one has 
$$
Int_{I_0}= \frac{V_n(x(t_0))^{\alpha p}}{\alpha p}\leq \frac{V_A^{\alpha p}}{\alpha p}\leq C_{4,p}\frac{A^p}p\leq C_{4,p}C_{3,p}\Vert d\Vert^p_{p,I_0}.
$$
For $I=I_j$, $1\leq j\leq J$ and $I_f$ we consider two cases, whether $\min_{I}V_n\geq \frac{v_A}2$ or not. 

In the first case, we rely on Eq.~\eqref{tech0} to obtain $Int_{I}\leq K(\frac{v_A}2)V_A^{\alpha p}\mid I\mid$. On the other hand, there exists $C_A>0$ such that $V_0(x)\geq C_A$ if $V_n(x)\geq \frac{v_A}2$. Therefore  $\mid I\mid$ is bounded by a constant times $\Vert V_0\Vert^p_{p,I}$
and one deduces the existence of a positive constant $C_{5,p}$ such that 
$$
Int_{I}\leq C_{5,p}\Vert d\Vert^p_{p,I_0}.
$$

Assume now that $\min_{I}V_n< \frac{v_A}2$. With no loss of generality, we can also assume that $Int_{I}>0$ otherwise we are done. If $\beta=\alpha(p-1)$ and the extremities of $I$ are $s$ and $t$, recall that 
$$
Int_{I}=\frac{V_n^{\beta+1}(x(t))-V_n^{\beta+1}(x(s))}{\beta+1},
$$ 
with $V_n(x(s))\geq v_A$. Then there exists $\tilde{s}<\tilde{t}$ in $(s,t)$ such that 
$$
V_n(x(\tilde{s}))=V_n(x(\tilde{t}))= v_A\hbox{ and }V_n(x(\cdot))\geq v_A\hbox{ on}(s,\tilde{s})\cup (\tilde{t},t).
$$
One deduces that $Int_{I}\leq \frac{\int_{\tilde{t}}^tV_n^\alpha\dot V}{\beta+1}$ and we are back to the first case.

Collecting all our estimates on the $Int_I$ yields the existence of a positive constant $C_{6,p}$ such that 
$$
\int_{V_{0,<}^{A,T}}V_n^{\alpha(p-1)}\dot V_n\leq C_{6,p}\Vert d\Vert^p_{p,V_{0,<}^{A,T}}.
$$
Gathering now Eq.~\eqref{=00} and \eqref{=0j} with the above estimate, we get the 
existence of a positive constant $C_{7,p}$ such that 
\beq\label{eq<}
\frac{V_0^p(x(T))}p+\int_{V_{0,<}^{A,T}}V_n^{\alpha(p-1)}\dot V_n+\frac{c_0}
4\int_{V_{0,<}^{A,T}}V_0^p\leq C_{7,p}\Vert d\Vert^p_{p,V_{0,<}^{A,T}}.
\eeq

Set $\tilde{c}=\frac{\min(c_n,c_0)}4$. By adding Eqs.~\eqref{eq>}, \eqref{eq=} and \eqref{eq<}, 
we get the existence of a positive constant $C_{8,p}$ such that 
\beq\label{eqToT}
\frac{V_0^p(x(T))}p+\frac{V_n^{\beta+1}(x(T))}{\beta+1}+\tilde{c}\int_0^TW^p\leq C_{8,p}\Vert d\Vert^p_{p,[0,T]},
\eeq
with possibly the term $\frac{V_n(x(T))^{\beta+1}}{\beta+1}$ not appearing if $I_f=\emptyset$.
In any case, by letting $T$ tends to infinity, we get Eq.~\eqref{th-fin}. As regards the convergence to the origin of any non trivial trajectory, first notice that $\liminf_{s\rightarrow \infty}x(s)=0$. Then, there is an increasing sequence of times $(t_l)$ tending to infinity so that 
$\lim_{l\rightarrow \infty}x(t_l)=0$. For $l\geq 0$, consider any time $T>t_l$ so that $x(t)$
remains in $V_{0,\leq}^{A}$ for $t\in [t_l,T]$. Multiplying Eq.~\ref{ineq2} by $V_0^{p-1}$ and integrating it over $[t_l,T]$, one gets that
$$
\frac{V_0^p(x(T))}p\leq \frac{V_0^p(x(t_l))}p+4l_0\int_{t_l}^{\infty}V_0^{p-1}\vert d\vert.
$$
The right-hand side tends to zero as $l$ tends to intinity. One deduces that for $l$ large enough, the trajectory remains in $V_{0,<}^{A}$ for $t\geq t_l$ and the above estimate is actually valid for every $t\geq t_l$. 

\begin{rem}\label{sat-ISS}
Eventhough we did not exhibit an ISS-Lyapunv function for $\dot x=J_nx-\frac{l_n}{\sigma_{+\infty}}e_n\sigma(k\omega(x)+d))$, the contents of Item $(S-\infty)$ in Proposition~\ref{th3} show that the above system is indeed ISS according to \cite[Theorem 2]{Sontag-ISS}
\end{rem}

{\bf Proof of Theorem~\ref{Th-p}. } In order to derive the theorem from Proposition~\ref{th3}, first remark the following: in the argument of Proposition~\ref{th3}, if the positive definite function $V_n$ is replaced by a positive definite function $Z$ veryfing Eqs.~\eqref{tech0} and \eqref{ineq1} for some positive constants $\widetilde{K}(C),\tilde{c_n},\tilde{d_n}$ and some $\tilde{\alpha}\in (0,1)$, then one obtains a proposition similar to Proposition~\ref{th3}
where, besides new constants in Eqs.~\eqref{th-inf} and \eqref{th-fin}, one replaces the positive definite function $W$ by a positive definite function $\widetilde{W}=\min(V_0(x),Z^{\tilde{\alpha}}(x))$. 

Recall that $\alpha=\frac{2(n-1)}{2n-1}$ was defined in Eq.~\eqref{der0}. For $\mu\in (1-\alpha,1]$, let $Z_\mu$ be the positive definite function equal to $V_n^{\mu}$. If $\dot Z_\mu$ denotes the derivative of $Z_\mu$ along non-trivial trajectories of the perturbed closed-loop system $\dot x=J_nx-e_nl_n\sigma(\omega(x))$, then $\dot Z_\mu=\mu V_n^{\mu-1}\dot V_n$ and one deduces at once the generalization of Eq.~\eqref{ineq1} only valid on the open set $V_{0,>}^A$,
$$
\dot Z_\mu\leq -\frac{\mu c_n}2 V_n^{\mu-1+\alpha}+\frac{4\mu l_n\vert d\vert}{V_n^{\mu-1}}\leq
-c_\mu Z_\mu^{\alpha_\mu}+l_\mu \vert d\vert,
$$
where $c_\mu,l_\mu$ are positive constants and $\alpha_\mu=\frac{\mu-1+\alpha}\mu$.
Since $Z_\mu^{\alpha_\mu}=V_n^{\mu-1+\alpha}$ one can use the preceding remark, one immediately deduces a proposition similar to Proposition~\ref{th3} for $W_\mu:=\min(V_0(x),V_n^{\mu-1+\alpha}(x))$. Furthermore, notice from Eq.~\eqref{wi-i} that, for $1\leq i\leq n$, there exists a positive constant $C_i$ so that $\vert x_i\vert^{\beta_{i-1}+1}\leq C_i W_i\leq C_i V_n$.
For $1\leq i\leq n$, first notice from Eq.~\eqref{wi-i} that there exists a positive constant $C_i$ so that $\vert x_i\vert^{\beta_{i-1}+1}\leq C_i W_i\leq C_i V_n$. After 
setting $\mu_i=1-\alpha+ \frac1{\beta_{i-1}+1}$, one gets that $\vert x_i\leq C'_iV_n^{\mu-1+\alpha}$
and then $\vert x_i\vert\leq C^{''}_i W_{\mu_i}$ for some positive constants $C'_i,C^{''}_i$. 
One deduces, for $1\leq i\leq n$, that the $L_p$-norm of $x_i$ is upper bounded by a constant times the $L_p$-norm of the internal disturbance $d$, and then the finite-gain property for the state feedback $u=\omega(x)$.

\section{$L_\infty$-stabilization in the presence of external disturbances} 
In this section, we focus on the $L_\infty$-stabilization of the perturbed system
\beq\label{sys-tot}
\dot x= J_nx+e_n\sigma(u+d)+E+d_ne_n,
\eeq
where $u,d,d_n\in\mathbb{R}$ and $E\in\mathbb{R}^{n-1}$ verifies $E^Te_n=0$. Here $d$ corresponds to an internal disturbance, $E$ to a mismatched external disturbance (i.e. misaligned with the input direction $e_n$) and $d_n$ stands for the matched external disturbance. We assume that both $d\in L_\infty(\mathbb{R}_+)$ and $E\in L_\infty(\mathbb{R}_+,\mathbb{R}^{n-1})$. As for $d_n$, we assume it belongs to the subspace $\Omega_\infty$ introduced in \cite{Saberi} and defined
$$
\Omega_\infty=\{f:\mathbb{R}_+\rightarrow\mathbb{R}, \ \hbox{ measurable such that } \sup_{t\geq 0}\vert\int_0^tf(s)ds\vert<\infty\}.
$$
For $f\in\Omega_\infty$ and $E=(d_1,\cdots,d_{n-1})\in L_\infty(\mathbb{R}_+,\mathbb{R}^{n-1})$, set 
\beq\label{norms}
N(f):=\lim_{t\rightarrow\infty}\sup_{t_2\geq t_1\geq t}\vert \int_{t_1}^{t_2}f(s)ds\vert,\quad
\Gamma(E):=\Vert E \Vert_{\infty}+\sum_{i=1}^{n-1}\Vert d_i\Vert_{\infty}^{\frac{2p_2}{p_{i+1}}}.
\eeq
We next provide a variant of the feedback $u=k(x)$ given by Theorem~\ref{Th-p} in order to $L_\infty$ stabilize the perturbed system~\eqref{sys-tot}. 

\begin{theorem}\label{th4} There exist positive constants $l_1,\cdots,l_n$ defining the function $\omega_n(\cdot)$ in Eq.~\eqref{u}, $A>0$ small enough so that Eq.~\eqref{estA} holds true, 
 and $k>0$ large enough, such that, if $\sigma$ an $S$-function, the dynamic feedback defined by $u=k\omega(x-ye_n)$ with $y(t)=\int_0^td_n(s)ds$, $t\geq 0$,  $L_\infty$-stabilizes the perturbed system~\eqref{sys-tot} in the following sense: there exists $C_\infty>0$ such that, for every $d\in L_\infty(\mathbb{R}_+)$, $E\in L_\infty(\mathbb{R}_+,\mathbb{R}^{n-1})$, $d_n\in\Omega_\infty$
and every trajectory of $\dot x=J_nx-\frac{l_n}{\sigma_{+\infty}}e_n\sigma(k\omega(x-ye_n)+d))+E+d_ne_n$, one has
\beq\label{est-per}
\limsup_{s\rightarrow\infty}W(x(s))\leq C_\infty\big(\Vert d\Vert_\infty+N(d_n)+\Gamma(E)\big).
\eeq
\end{theorem}

{\bf Proof. } Set $E=(d_1,\cdots,d_{n-1})^T$ gathering the $n-1$ mismatched scalar external disturbances. First of all, note that $y(\cdot)$ is an $L_\infty$-function since $d_n\in \Omega_\infty$. By performing the change of variable $X=x-ye_n$, the perturbed system
$\dot x=J_nx-\frac{l_n}{\sigma_{+\infty}}e_n\sigma(k\omega(x-ye_n)+d))+E+d_ne_n$ reduces 
$\dot X=J_nX-\frac{l_n}{\sigma_{+\infty}}e_n\sigma(k\omega(X)+d))+F$ with a mismatched disturbance $F=(d_1,\cdots,y+d_{n-1})^T$. It is therefore enough to prove the theorem in the case $d_n=0$ and thus $y=0$.

We essentially follow the lines of the proof of Proposition~\ref{th3}. For that purpose, one needs to modify inequalities~\eqref{ineq1}, \eqref{ineq2} so as to take into account the mismatched disturbance $E$. Since the Lyapunov function $V_0$ is quadratic, it is immediate to get an inequality extending Eq.~\eqref{ineq2} where the term $\min(1,\vert d\vert)$
is replaced by $\min(1,\vert d\vert)+\Vert E\Vert$ by possibly changing the constants $c_0,l_0$. 

As concerns the modification of Eq.~\eqref{ineq1}, the main ingredient consists of the following extension of Eq.~\eqref{der0}  in the presence of  the mismatched disturbance $E$, which is proved in Appendix: there exist  positive constants $l_1,\cdots,l_n$ defining the function $\omega_n(\cdot)$ in Eq.~\eqref{u} so that the time derivative of $V_n$ along non trivial trajectories of $\dot x=J_nx+e_nu+E$, where $E^Te_n=0$, can be upper bounded as next,
\beq\label{der-mis}
\dot V_n\leq -C_1V_n^{\alpha}(x)+\omega_n(x)(u+l_nsign(\omega_n(x))+C_2\sum_{i=1}^{n-1}\vert d_i\vert^{\frac{2p_2}{p_{i+1}}},
\eeq
where $C_1,C_2$ are positive constants. It is then immediate to get Eq.~\eqref{tech0} from the argument given for Eq.~\eqref{der-mis}. From that, we simply reproduce the same arguments given to obtain Eq.~\eqref{ineq1} to derive its generalization corresponding to the presence of the mismatched disturbance $E$: one replaces the term $4l_n\vert d\vert$ by $4L_n(\vert d\vert+\sum_{i=1}^{n-1}\vert d_i\vert^{\frac{2p_2}{p_{i+1}}})$ for some positive constant $L_n$. The proof of Theorem~\ref{th4} then proceeds as that of Item $(S-\infty)$ in Theorem~\ref{th3} and one gets Theorem~\ref{th4}.

\begin{rem}\label{notP}
One should notice the solution proposed in Theorem~\ref{th4} for the $L_\infty$-stabilisation of
the perturbed system \eqref{sys-tot}, as well as that given in Theorems $2$ and $3$ in \cite{Saberi} present a possible restrictive feature when the matched perturbation $d_n$ is not zero because, for all of them, the proposed feedbacks depend on $d_n$.
\end{rem}

\section{Appendix} 
\subsection{Proof of Eqs.~\eqref{ineq1} and \eqref{ineq2}}
We next provide an argument for Eq.~\eqref{ineq1}. Consider a trajectory of $\dot x=J_nx-l_ne_n\sigma(k\omega(x)+d)$ lying in $V_{0,>}^A$. Then one has 
$$
\dot x=J_nx-l_ne_n sign(k\omega_n(x)+d)-l_ne_n\big(\sigma(k\omega_n(x)+d)-sign(k\omega_n(x)+d)\big).
$$
Set $\xi(t)= k\omega_n(x(t))+d(t)$. Using Eq.~\eqref{der1}, one deduces that
$$
\dot V_n\leq -c_nV_n^{\alpha}(x(t))+2(1+\frac1k)l_n\vert d(t)\vert+\frac{l_n}k\vert \xi(t)\vert \mid \sigma(\xi(t))-sign(\xi(t))\mid.
$$
If $\vert k\omega_n(x(t))+d(t)\vert\geq 1$, then, by using Eq.~\eqref{esti-SF}
$$
\vert \xi(t)\vert \mid \sigma(\xi(t))-sign(\xi(t))\mid\leq \frac{C_\sigma\vert \xi(t)\vert }{1+\vert\xi(t)\vert}\leq C_\sigma.
$$
Otherwise, $\dot V_n\leq -c_nV_n^{\alpha}(x(t))+2(1+\frac1k)l_n\vert d(t)\vert+\frac{2l_n}k$, 
which implies that one always has that
\begin{equation}\label{partial1}
\dot V_n\leq -c_nV_n^{\alpha}(x(t))+\frac{(2+C_\sigma)l_n}k+2(1+\frac2k)l_n\vert d(t)\vert.
\end{equation}
Using the fact that the trajectories lies in $V_{0,>}^A$, one finally deduces that
$$
\dot V_n\leq -\frac{c_n}2V_n^{\alpha}(x(t))-\frac{c_n}2v_A^\alpha+\frac{(2+C_\sigma)l_n}k+2(1+\frac2k)l_n\vert d(t)\vert.
$$
By taking $k\geq \max(2,\frac{2(2+C_\sigma)l_n}{c_nv_A^\alpha})$, one derives Eq.~\eqref{ineq1}. 

We now turn to a proof for Eq.~\eqref{ineq2}. Set $\underline{\rho}:=\min_{\mid s\mid\leq 1}\frac{\sigma(s)}s>0$ and $\bar{\rho}:=\max_{\mid s\mid\leq 1}\frac{\sigma(s)}s$. Consider a trajectory of $\dot x=J_nx-l_ne_n\sigma(k\omega(x)+d)$ lying in $V_{0,\leq}^A$. Then one has 
$$
\dot x=(J_n-r(t)e_nK^T)x-l_ne_n\big(\sigma(\omega_0(x)+d)-\sigma(\omega_0(x))\big),
$$
where $r(t)=\frac{\sigma(\omega_0(x(t)))}{\omega_0(x(t))}$ and $r(t)\in [\underline{\rho},\bar{\rho}]$. We can now use Item $(i)$ of Definition~\ref{def-sat}, apply Eq.~\eqref{der2} and conclude.
\subsection{Proof of Eq.~\eqref{der-mis}}
The argument actually consists of following the steps of the original proof of Eq.~\eqref{der0} as elaborated by Hong in \cite{Hong} while incorporating the external disturbances $d_1,\cdots,d_{n-1}$ and handling their effect. 

To this end, we need to recall several technical data used in  \cite{Hong} and in particular to precise the notion of homogeneity mentioned when the Lyapunov function $V_n$ was first considered in Eq.~\eqref{V_n}. For $1\leq i\leq n$ and $\varepsilon>0$, let $\delta^{\bar{p}_i}_\varepsilon$ be the family of dilations defined on $\mathbb{R}^i$ by $\delta^{\bar{p}_i}_\varepsilon(x)=(\varepsilon^{p_1}x_1,\cdots,\varepsilon^{p_i}x_i)$ where $x=(x_1,\cdots,x_i)\in \mathbb{R}^i$, $\bar{p}_i=(p_1,\cdots,p_i)$ is defined in Eq.~\eqref{param0}. A function $V:\mathbb{R}^i\rightarrow\mathbb{R}$ is said to be homogeneous of degree $\alpha>0$ (with respect to the family of dilations $\delta^{\bar{p}_i}_\varepsilon$) if $V(\varepsilon^{p_1}x_1,\cdots,\varepsilon^{p_i}x_i)=\varepsilon^{\alpha}V(x)$ for every $x\in \mathbb{R}^i$. 

For $1\leq i\leq n$ define the positive definite function $V_i:\mathbb{R}^i\rightarrow\mathbb{R}_+$ as $V_i(x)=\sum_{j=1}^iW_j(x_1,\cdots,x_j)$ and, for $1\leq i\leq n-1$,  the constants 
$$
\alpha_i=\frac{2p_2}{1+p_2-p_i},\quad \eta_i=\frac{2p_2}{p_{i+1}}.
$$
Note that $\frac1{\alpha_i}+\frac1{\eta_i}=1$ for $1\leq i\leq n-1$.
As proved in \cite{Hong}, one has that, for $1\leq i\leq n$, $W_i$ and $V_i$ are homogeneous of degree $1+p_2$ and, along non trivial trajectories of the unperturbed system $\dot x=J_nx+e_nu$, the time derivative $\dot V_i$ of $V_i$ is homogeneous of degree $2p_2$.

For $1\leq i\leq n-1$, we prove by induction that there exist positive constants $l_1,\cdots, l_{n-1}$  so that 
\beq\label{cru-fin}
\dot V_i\leq -\sum_{j=1}^i\frac{l_j}2\vert \omega_j\vert^{\alpha_j}+\omega_i(x_{i+1}-\nu_i)+C_i\sum_{j=1}^i\vert d_j\vert^{\eta_j}.
\eeq 

We start the induction at $i=1$ and get, for any choice of positive $l_1$, 
$$
\dot V_1=\lf x_1\rr^{p_2}(x_2+d_1)\leq -l_1\vert \omega_1(x_1)\vert^{\alpha_1}
+ \omega_1(x_1)(x_2-v_1)+\omega_1(x_1) d_1.
$$
By using Young's inequality, one gets $\vert \omega_1(x_1) d_2\vert\leq \frac{l_1}2\vert \omega_1(x_1)\vert^{\alpha_1}+c_1\vert d_1\vert^{\eta_1}$, for some positive constant $c_1$, and hence Eq.~\eqref{cru-fin} for $i=1$.

Assume we have established Eq.~\eqref{cru-fin} for $i-1$ with $i\leq n-1$ and some positive constants $l_1,\cdots,l_{i-1}$. Then one gets 
$$
\begin{array}{ccc}
\dot V_i&=&\dot V_{i-1}+\sum_{j=1}^{i-1}\frac{\partial W_i}{\partial x_j}(x_{j+1}+d_j)+\omega_i(x_{i+1}-\nu_i)+\omega_i\nu_i+\omega_id_i,\\
&\leq&-\sum_{j=1}^i\frac{l_j}2\vert \omega_j\vert^{\alpha_j}+\omega_i(x_{i+1}-\nu_i)+C_{i-1}\sum_{j=1}^{i-1}\vert d_j\vert^{\eta_j}+V_i^0,
\end{array}
$$
where $l_i>0$ will be chosen below and 
$$
V_j^0=-\frac{l_i}2\vert \omega_i\vert^{\alpha_i}+\sum_{j=1}^{i-1}\frac{\partial W_i}{\partial x_j}(x_{j+1}+d_j)+\omega_{i-1}(x_i-\nu_{i-1})+\omega_id_i,
$$
By applying Young's inequality to $\vert \frac{\partial W_i}{\partial x_j}d_j\vert$ and $\vert\omega_id_i\vert$, one deduces that 
$
V_j^0\leq V_j^1+c_i\vert d_i\vert^{\eta_i}$,
where $$V_j^1=-\frac{l_i}4\vert \omega_i\vert^{\alpha_i}+\sum_{j=1}^{i-1}\frac{\partial W_i}{\partial x_j}x_{j+1}+\sum_{j=1}^{i-1}\frac1{\alpha_j}\vert\frac{\partial W_i}{\partial x_j}\vert^{\alpha_j}+\omega_{i-1}(x_i-\nu_{i-1}).$$
The last step of the reasoning consists of showing that $l_i>0$ can be chosen large enough so that $V_j^1\leq 0$. This is done by first noticing that $V_j^1$ is homogeneous of degree $2p_2$ and by checking that the homogeneity argument provided at the end of page $234$ and the top of page $235$ of \cite{Hong} exactly applies to the present situation. That concludes the induction step and the proof of Eq~\eqref{cru-fin}.

Again by following the end of the argument in the top of page $235$ of \cite{Hong}, one deduces 
Eq.~\eqref{der-mis} from Eq~\eqref{cru-fin} since there is no external disturbance for the dynamics of $x_n$.

\end{document}